\def\lax    {\ifmmode{_<\atop^{\sim}}\else{${_<\atop^{\sim}}$}\fi}
\def\gax    {\ifmmode{_>\atop^{\sim}}\else{${_>\atop^{\sim}}$}\fi}
\newbox\grsign      \setbox\grsign=\hbox{$>$}
\newdimen\grdimen   \grdimen=\ht\grsign
\newbox\simgreatbox \setbox\simgreatbox=\hbox{\raise.5ex\hbox{$>$}\llap
                        {\lower.5ex\hbox{$\sim$}}}\ht1=\grdimen\dp1=0pt
\newbox\simlessbox  \setbox\simlessbox =\hbox{\raise.5ex\hbox{$<$}\llap
                        {\lower.5ex\hbox{$\sim$}}}\ht2=\grdimen\dp2=0pt
\def\simgreat{\mathrel{\copy\simgreatbox}}
\def\simless {\mathrel{\copy\simlessbox }}

\newbox\grsign \setbox\grsign=\hbox{$>$} \newdimen\grdimen \grdimen=\ht\grsign
\newbox\laxbox \newbox\gaxbox
\setbox\gaxbox=\hbox{\raise.5ex\hbox{$>$}\llap
     {\lower.5ex\hbox{$\sim$}}}\ht1=\grdimen\dp1=0pt
\setbox\laxbox=\hbox{\raise.5ex\hbox{$<$}\llap
     {\lower.5ex\hbox{$\sim$}}}\ht2=\grdimen\dp2=0pt
\def\gax{\mathrel{\copy\gaxbox}}
\def\lax{\mathrel{\copy\laxbox}}
\def\boxit#1    {\vbox{\hrule\hbox{\vrule\kern3pt
                  \vbox{\kern3pt#1\kern3pt}\kern3pt\vrule}\hrule}}
\def\h      {\ifmmode{^{\rm h}}\else{$^{\rm h}$}\fi}
\def\m      {\ifmmode{^{\rm m}}\else{$^{\rm m}$}\fi}
\def\s      {\ifmmode{^{\rm s}}\else{$^{\rm s}$}\fi}
\def\decas    {\ifmmode{{\rlap.}{''}}\else{${\rlap.}{''}$}\fi}
\def\mum     {\ifmmode{\mu{\rm m}}\else{$\mu{\rm m}$}\fi}
\def\s      {\ifmmode{^{\rm s}}\else{$^{\rm s}$}\fi}
\def\deg      {\ifmmode{^{\circ}}\else{$^{\circ}$}\fi}
\def\as     {\ifmmode {\rlap.}$\,$''$\,$\! \else ${\rlap.}$\,$''$\,$\!$\fi}
\def\decsec  {\ifmmode {\rlap.}$\,$^{s}$\,$\! \else ${\rlap.}$\,$^{s}$\,$\!$\fi}\def\decs  {\ifmmode {\rlap.}$\,$^{s}$\,$\! \else ${\rlap.}$\,$^{s}$\,$\!$\fi}

\def\kms    {\ifmmode{{\rm km~s}^{-1}}\else{km~s$^{-1}$}\fi}

\def\ccm    {cm$^{-3}$}
\def\scm    {cm$^{-2}$}

\def\Lsun   {$L_{\odot}$}

\def\Msun   {$M_{\odot}$}
\def\Mspy   {\ifmmode {M_{\odot} {\rm yr}^{-1}} \else $M_{\odot}$~yr$^{-1}$\fi}
\def\Mdot   {\ifmmode {\dot M} \else $\dot M$\fi}
\def\mhd    {\ifmmode {n_{{\rm H}_2}} \else $n_{{\rm H}_2}$\fi}
\def\mhcd   {\ifmmode {N_{{\rm H}_2}} \else $N_{{\rm H}_2}$\fi}

\def\El      {\ifmmode{E_{\ell}}\else{$E_{\ell}$}\fi}
\def\beam    {\ifmmode{\theta_{\rm B}}\else{$\theta_{\rm B}$}\fi}
\def\mjyb   {\ifmmode {{\rm mJy~beam}^{-1}} \else{mJy~beam$^{-1}$}\fi}
\def\mujyb   {\ifmmode {\mu{\rm Jy~beam}^{-1}} \else{$\mu$Jy~beam$^{-1}$}\fi}
\def\Trot   {\ifmmode{T_{\rm rot}}\else$T_{\rm rot}$\fi}

\def\Teff   {\ifmmode{T_{\rm eff}}\else$T_{\rm eff}$\fi}

\def\Tkin   {$T_{\rm kin}$}

\def\ITRS   {\ifmmode{\smallint {\rm T}_{R}^{*}dv}\else{$\smallint
{\rm T}_{R}^{*}dv$}\fi}
\def\ITRS   {\ifmmode{\smallint {\rm T}_{R}^{*}dv}\else{$\smallint
{\rm T}_{R}^{*}dv$}\fi}
\def\ITAS   {\ifmmode{\smallint {\rm T}_{A}^{*}dv}\else{$\smallint
{\rm T}_{A}^{*}dv$}\fi}

\def\hh         {H$_2$}

\def\hhco       {H$_2$CO}

\def\nhhh       {NH$_3$}

\documentclass{iaus}
\usepackage{graphicx}
\usepackage{ulem}
\usepackage{upmath}
\usepackage{dcolumn}
\newcolumntype{d}{D{.}{.}{-1}}
\newcommand{\dlabel}[1]{\multicolumn{1}{c}{\mbox{#1}}}
\title[Initial Conditions -- IRDCs] 
{Initial Conditions for Massive Star Birth -- Infrared Dark Clouds}

\author[Menten, Pillai, \&\ Wyrowski]   
{K. M. Menten, T. Pillai, \and F. Wyrowski}
\affiliation{Max-Planck-Institut f\"ur Radioastronomie, Auf dem H\"ugel 69, D-53121 Bonn,
Germany}

\pubyear{2005}
\volume{227}  
\date{?? and in revised form ??}
\jname{Massive Star Birth: A Crossroads of Astrophysics }
\editors{R. Cesaroni, E. Churchwell, M. Felli, \& C.M. Walmsley, eds.}
\begin{document}

\maketitle

\begin{abstract}
We summarize the properties of Infrared Dark Clouds, massive, dense,
and cool aggregations of interstellar gas and dust that are found
througout the Galaxy in projection against the strong mid-infrared
background. We describe their distribution and give an overview of
their physical properties and chemistry. These objects appear to be
the progenitors of high-mass stars and star clusters, but seem to be
largely devoid of star formation, which however is taking place in
localized spots.

%
%

\end{abstract}

\firstsection 
\section{Introduction}
Observations of dust and molecules
provide almost all of the accessible information on
deeply embedded high-mass (pre-)protostars, whose emission is frequently
not detected even at mid-infrared wavelengths. As evidenced by, a.o., many
contributions to this symposium, much current effort is
expended on extensive surveys of unequivocal signposts
of high-mass star formation, such as sources with tell-tale far-infrared spectral energy
distributions and hot, dense molecular cores highlighted by maser
emission in the methanol and water molecules.

All the regions traced by these signposts are containing already
formed (proto)stars and only recently have clouds with the potential
of forming high-mass stars and/or clusters, but still yet largely
devoid of stellar objects, been identified: Infrared Dark Clouds,
whose observational status we shall summarize here.

\section{Infrared Dark Clouds}\label{IRDCs}
First recognized in mid-infrared images from the Infrared Space
Observatory (ISO) and Midcourse Space Experiment (MSX) satellites
Infrared Dark Clouds (IRDCs) appear in silhouette against the Galactic
mid-infrared (MIR) background, frequently in filamentary shapes.
Perault et al. (1996) report that ISOGAL\footnote{ISOGAL is a 7 and
$15 \mu$m survey with ISOCAM (the ISO 3 -- $20 \mu$m camera) of
$\sim 12\deg$ in the Galactic Plane interior to $|l| = 45\deg$.}
images show ``unexpectedly, a number of regions which are optically
thick at $15 \mu$m which are likely due to absorption" and ``a
convincing correlation with a depletion in $2 \mu$m source
counts". They estimate $A_V > 25$ and put forward ``that these would
be good candidates for precursors of star formation sites."
Even before IRDCs became generally known as a distinct class of
objects, Lis \& Menten (1998) found absorption in the 45 $\mu$m ISO
LWS detector band against the MIR background and emission in the 173
$\mu$m band toward M0.25+0.11, the low Galactic longitude end of the
Galactic center ``dust ridge', a string of submillimeter (submm)
condensations found by Lis \& Carlstrom (1994) which terminates with the prominent
Sgr B2 star-forming region at its high longitude end. M0.25+0.11 was
studied, even earlier, in detail by Lis et al. (1994), who found very little, if any
(a weak H$_2$O maser) signs of star formation in it.  Lis \& Menten
performed grey body spectral fits to the far-infrared data combined
with their previous 350 -- 800 $\mu$m submillimeter measurements and
obtained a low temperature, $\sim 18$ K, for the bulk of the dust in
M0.25+0.11's core.  In addition, they found that the grain emissivity
is a very steep function of frequency ($\nu^{2.8}$; see
\S\ref{cdandt}).   Lis
\& Menten derived a gigantic mass of $1\times10^6$ \Msun\ for this
object, comparable to the core of the ``mini starburst region'' Sgr
B2.

The first extensive dataset on IRDCs was compiled with the
SPIRIT III infrared telescope aboard the Midcourse Space Experiment
(MSX; see Price 1995), which
surveyed the whole Galactic plane in a $b = \pm 5\deg$ wide strip
(Price et al. 2001)
in four MIR spectral bands between 6 and 25 $\mu$m at a spatial
resolution of $\sim18\as3$.
In an initial census of a $\sim180\deg$ long strip
of the Galactic plane (between $269\deg < l < 91\deg, b = \pm 0.5\deg$),
Egan et al. (1998) find $\sim2000$
``compact objects seen in absorption against bright mid-infrared emission
from the Galactic plane.
Examination of MSX and IRAS images of these objects reveal that they
are dark from 7 to 100 $\mu$m.''

The IRDCs are best identified in the $8 \mu$m MSX ``A'' band,
because, first, the 7.7 and 8.6~$ \mu$m PAH features associated with
star-forming regions contribute to a brighter background emission and, second, the
MSX A band is more sensitive than the satellite's other bands.

M0.25+0.11 and the other condensations of the Galactic center
dust ridge also appear
conspicuously in absorption on an $8 \mu$m MSX image presented by Egan et
al. 1998.

Hennebelle et al. (2001) in a systematic analysis of the ISOGAL images
extracted a total of $\sim 450$ IRDCs, for which they derive $15 \mu$m opacities of 1 to 4.

The four newly-detected
massive and dense cold cores identified by Garay et al.\ (2004; see
also his contribution to these proceedings)  also represent an interesting
sample of IRDCs. These objects are not mid-IR but mm-contiumm selected:
The 1.2~mm dust emission reveals massive
($M>400 M_{\odot}$) and cold ($T<16$~K) cores.

How do IRDCs compare to the Orion Molecular Cloud I (OMC-1), probably
the best-studied high-mass molecular cloud/star-forming region
complex?  In Fig.\ 1, we show the 1.2~mm dust continuum emission
from OMC-1 mapped with MAMBO\footnote{The \textbf{MA}x-Planck  \textbf{M}illimeter
\textbf{BO}lometer array is operated at the IRAM 30m telescope on
Pico Veleta, Spain.} and the 850$\mu$m
SCUBA\footnote{The \textbf{S}ubmillimeter \textbf{C}ommon \textbf{U}ser
\textbf{B}olometer \textbf{A}rray is operated at the 15 m James-Cleck-Maxwell
Telescope on Mauna Kea, Hawaii.} dust
continuum map of IRDC G11.11-0.12 (Johnstone at al. 2003).  At a
distance of 3.6~kpc (see \S\ref{cdandt}) G11.11-0.12 has a
remarkable resemblance with the integral-shaped Orion filament,
both, in structure and in dimensions.  The
bright MIR emission along the Galactic plane favours the
identification of a massive cold cloud as infrared dark, which the
OMC-1 region is not due to the absence of a MIR background caused by
its location outside of the Galactic plane. One glaring difference
though exists between the two maps: the prominent maximum in OMC-1,
marking the very active BN/KL high-mass stars-forming region
(see \S\ref{starformation}).
\label{fig:1}
\begin{figure}
\begin{center}
 \includegraphics[height=0.9\linewidth,angle=-90]{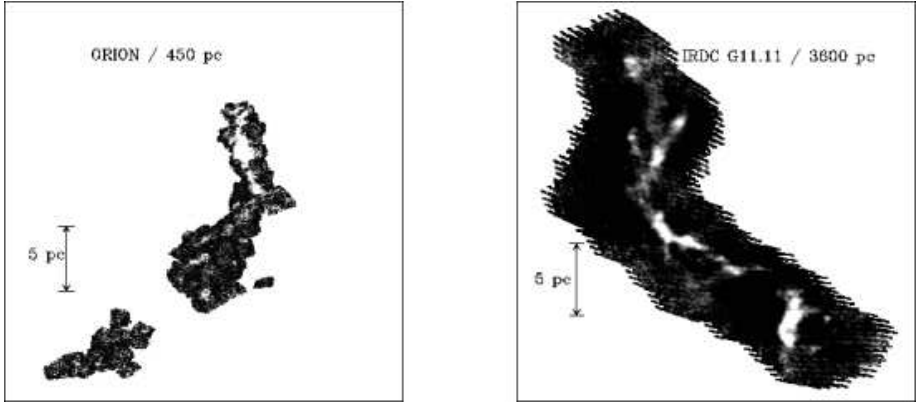}
\end{center}
 \caption{\textit{Left panel:} 1.2~mm dust continuum map of the Orion Molecular Cloud 1
 (courtesy T. Stanke). \textit{Right panel:} $850~\mu$m map of G11.11-0.12 (Johnstone et al.2003)}
\end{figure}

\section{Galactic distribution and distances}\label{distance}
Using a standard Galactic rotation curve, Carey et al. (1998) determined kinematic distances
for  some of the IRDCs in their sample. They obtain distances between 2.2 and 4.8 kpc,
proving without a doubt that the clouds are not local. Their distances agree with those
of HII regions in their vicinity.

Recently, Simon et al. (2004) prepared a catalogue of $\sim$~380 IRDCs
identified from the MSX survey, based on the morphological correlation
of MIR extinction and $^{13}$CO emission, as observed in the BU-FCRAO
Galactic ring survey (GRS) of the inner Milky Way. They find that the
majority of the dark clouds are concentrated in the Galactic ring at a
galactocentric radius of 5~kpc. The kinematic distances derived range
from 2 -- 9~kpc. In Table 1, we list the source parameters of all the
IRDCs for which data at wavelengths other than the MIR have been
collected.

\begin{table}
\begin{center}
\caption{List of all IRDCs with (sub)mm-wavelength data}

\vspace*{2mm}
\begin{tabular}{llldddd}
Name   &        R.A.  &     Decl.      & \dlabel{$V_{\rm LSR}$}   &     \dlabel{ d }   & \dlabel{$N({\rm H+H_2})$} &   \dlabel{\Tkin} \\
\hline
       & (J2000.0)   &   (J2000.0)  &  \dlabel{(km$\,$s$^{-1}$)}&   \dlabel{  (kpc)}    &  \dlabel{($10^{22}~{\rm cm^{-2}}$)} & \dlabel{(K)}  \\
\hline

DF+04.36-0.06 $^t$    & 17:55:53.07 & $-$25:13:18.7 &11.4   & 3.5          &  -   &  -     \\
DF+09.86-0.04 $^t$    & 18:07:37.22 & $-$20:25:54.5 &17.7   & 2.8          &  6.1  &  10.0     \\
DF+15.05+0.09 $^t$    & 18:17:37.87 & $-$15:48:59.9 &29.9   & 3.1          &  12.6 &  -     \\
DF+18.56-0.15 $^t$    & 18:25:19.52 & $-$12:49:57.0 &50.5   & 4.0          &  -   &  8.0     \\
DF+18.79-0.03 $^t$    & 18:25:19.84 & $-$12:34:23.1 &-     & 3.6          &  -   &  -     \\
DF+25.90-0.17 $^t$    & 18:39:10.13 & $-$06:19:58.8 &-     & 5.5          &  -   &  -     \\
DF+30.23-0.20 $^t$    & 18:47:13.16 & $-$02:29:44.7 &104.7  & 6.7          &  11.1 &  8.0     \\
DF+30.31-0.28 $^t$    & 18:47:39.03 & $-$02:27:39.8 &-     & 6.3          &  -   &  -     \\
DF+30.36+0.11 $^t$    & 18:46:21.16 & $-$02:14:19.0 & 96.1  & 5.9          &  -   &  -     \\
DF+30.36-0.27 $^t$    & 18:47:42.37 & $-$02:24:43.2 &-     & 6.9          &  -   &  -     \\
DF+31.03+0.27 $^t$    & 18:47:00.39 & $-$01:34:10.0 &77.8   & 4.9          & 11.1  &  10.0    \\
DF+36.95+0.22 $^t$    & 18:57:59.51 &   +03:40:33.3 &-     & 5.0          &  -         &  -     \\
DF+51.47+0.00 $^t$    & 19:26:12.74 &   +16:26:12.6 & 54.7  & 5.3          &  7.7      &  10.0    \\
G353.85+0.23 P1 $^c$  & 17:29:16.5  & $-$34:00:06   &-     & -           &  -       &    -   \\
G353.51-0.33 P1 $^c$  & 17:30:26.0  & $-$34:41:48   &-     & -           &  -       &    -  \\
G357.51+0.33 P1  $^c$ & 17:40:49.9  & $-$31:14:50   &-     & -           &  -      &     -  \\
G10.74-0.13 P1  $^c$  & 18:09:45.9  & $-$19:42:04   &-     & -           &  -      &     -  \\
G11.11-0.12P1$^p$     & 18:10:29.27 & $-$19:22:40.3 & 29.2  & 3.6          & 1.7       & 13.5 \\
G19.30+0.07 $^p$      & 18:25:56.78 & $-$12:04:25.0 & 26.3  & 2.2          &  -          & 18.5 \\
G24.72-0.75 $^p$      & 18:36:21.07 & $-$07:41:37.7 & 56.4  & 3.6          &  -         & 20.3 \\
G24.63+0.15 $^p$      & 18:35:40.44 & $-$07:18:42.3 & 54.2  & 3.6          &  -         & 14.4 \\
G28.34+0.06P1$^p$     & 18:42:50.9  & $-$04:03:14   & 78.4  & 4.8          & 3.3        & 16.6 \\
G28.34+0.06P2$^p$     & 18:42:52.4  & $-$03:59:54   & 78.4  & 4.8          & 9.3        & 16.0 \\
G33.71-0.01$^p$       & 18:52:53.81 &   +00:41:06.4 & 104.2 & 7.2          & -          & 17.2 \\
G79.27+0.38 $^p$      & 20:31:59.61 &   +40:18:26.4 & 1.2   & 1.0          &8.3        & 11.7 \\
G79.34+0.33 $^p$      & 20:32:26.20 &   +40:19:40.9 & 0.1   & 1.0          &8.8        & 14.6 \\
G81.50+0.14 $^p$      & 20:40:08.29 &   +41:56:26.4 & 8.7   & 1.3          & -          & 16.6 \\

\hline

\end{tabular}
\end{center}
Notes: Columns are name, right ascension, declination, LSR velocity, distance, total
column density, and kinetic temperature. In souces with a distance ambiguity,
the near distance was chosen. $^t$ refers to sources from Teyssier et al. 2002, $^c$
from Carey et al. (2000), 'P1/P2' refers to the brightest submm peaks
of Carey et al. (2000), $^p$ means that the position of the brightest ammonia  peak is given rather
than the submm peak position (see Pillai et al. 2005a).
\label{source parameters}
\end{table}

The cloud sizes as reported by Carey et al. (1998) are from 0.4 -- 15~pc while
Teyssier at al. (2002) report structures of filaments down to sizes $\le1$~pc.

\section{Physical parameters or IRDCs}
\subsection{Density, column density and temperature}\label{cdandt}
Based on observations of the formaldehyde molecule (\hhco), Carey et al. (1998)
argue that IRDCs objects are dense ($n > 10^5$ cm$^{-3}$), cold ($T < 20$ K) cores,
apparently without surrounding envelopes. However,
later work by Teyssier at al. on a different IRDC sample report that large
field maps obtained with the 4-m Nanten telescope in the $^{13}$CO J$=1-0$
line, which probes relatively low densities,
(Zagury et al., unpublished data) indicate that at least these IRDCs
may indeed have lower density envelopes. We have confirmed this for the Carey et
al. sample with hitherto unpublished $^{13}$CO $J = 1-0$ maps retrieved from
the GRS.

Carey et al. (1998) conclude that the IRDCs they have studied
have extinctions in excess of 2 mag at 8 $\mu$m.
Using the infrared  (visual through $30 \mu$m)
extinction law Lutz et al. (1996) derived for the Galactic center,
this indicates visual extinctions of $> 30$ mag.
How representative Lutz et al.'s law is for other lines of sight is unknown.
Certainly it does lack the  pronounced minimum for standard
graphite-silicate mixes in the  4 -- 8 $\mu$m range predicted by Draine \&\ Lee (1984),
which seems well-established by observations toward various lines-of-sight (see references
therein).

Actually, IRDC opacities at two wavelengths (e.g. 7 and 15 $\mu$m) can be used as
a check on the extinction curve and its possible variation with different lines
of sight; see Teyssier et al. 2002. These authors find a (marginally) lower
7 to 15 $\mu$m opacity ratio for clouds located away from the Galactic center compared to
clouds that appear in the Galactic center direction. For the latter value they derive
$\sim 1$, which is consistent with Lutz et al.'s law.

 Using the relation given by
Bohlin et al. (1978),
assuming that all the hydrogen is in molecular form and a ``standard"
ratio of total to selective extinction
of 3.1, hydrogen column densities in excess of $3\times10^{22}$ \ccm\ is derived.

An independent \hh\ column density estimate can be obtained from observations of
(sub)millimeter dust emission, which, in addition, also allow
determination of the cloud (gas+dust) mass (see, e.g.,  Mezger et al. 1987, 1990; Lis \&\ Menten
1998).
At (sub)millimeter wavelengths the dust emission is generally optically thin over most
%
of the volume of an interstellar cloud. Thus, at wavelength
$\lambda$  the measured flux density,  $S_\lambda$, is given
by $\int B_\lambda (T_{\rm D})(1 - e^{-{\tau}_{\lambda}}){\rm d}\Omega$,
where  $T_{\rm D}$ is the  temperature of the dust, ${\tau}_{\lambda}$
its optical depth and $B_\lambda$ is the Planck black body brightness.
The integration is either over the beam solid angle for a point-like source
or over the source's angular extent, if the latter is extended.
${\tau}_{\lambda}$ is proportional to the hydrogen column density, $N({\rm H}_2)$,
and the dust absorption cross section per hydrogen atom $\sigma_\lambda$,
which itself is assumed to be proportional to $\lambda^{-\beta}$.

Using all of the above, one finds that the  \hh\ column density  is related to
$S_\lambda$ as
$N({\rm H}_2) \propto S_\lambda \lambda^{3+\beta} (e^{hc/\lambda T_{\rm D}} - 1)$.
The cloud mass, $M$,
is proportional to $N({\rm H}_2) D^2$, where $D$ is the cloud's distance.

Carey et al. (2000) imaged a sample of 8 IRDCs in $450$ and $850 \mu$m dust
emission using SCUBA. Since it impossible to
determine, both, $T_{\rm D}$ and $\beta$ with just two data points, Carey et al.
calculated dust color temperatures for three different values of
$\beta$, 1.5, 1.75, and 2 and note that higher $\beta$-values (meaning
lower temperatures) yield
a better fit to the low temperatures Carey et al. (1998) obtained from \hhco\
observations. The (high) column densities implied by choosing $\beta = 2$ are
around $5\times 10^{22}$ \scm\ for 4 sources of their sample and
around  $13\times 10^{22}$ \scm\ for three.
Three of the cores corresponding to the brightest submm peaks have masses around 100 \Msun, two other have 400 and 1200
\Msun, respectively. Two clouds in the Cygnus region have masses around 40 \Msun,
but we note that distance estimates for that region are very uncertain and
probably a short distance (1 kpc used by Carey et al. 1998) was
chosen for the  latter calculations.
For $\beta = 1.75$, all these values are
to be reduced by a factor of 2.

As reported above (see \S\ref{IRDCs}), for M0.25
Lis \& Menten derive a very high value of $\beta$ of 2.8.
They take that to indicate the presence of dust grains covered with thick ice mantles.

The values for the high densities and low temperatures deduced by Egan et al. (1998)
are confirmed by Carey et al. (1998), who made, for 10 IRDCs,
millimeter-wavelength observations of \hhco, which is a well-established density probe
(Mangum \& Wootten 1993; Mundy et al. 1987). Since they observed several
transition, they were able, using a Large Velocity Gradient method, to determine
temperatures and abundances. Unfortunately, these authors do not
discuss their results source by source, but only give general statements.

Leurini in her dissertation (Bonn University; see also Leurini et al. 2004) has shown
that methanol (CH$_3$OH) is a highly useful interstellar density and temperature probe.
Consequently, she conducted observations of IRDCs in a selected series of
lines of that molecule, which she showed to be overabundant in these sources
(see \S\ref{chemistry})
and, thus, easy to detect.  Leurini also corroborates the high
densities ($\sim10^5$ -- $10^6$ \ccm) indicated by the \hhco\ data.
However, she only observed positions of submillimeter emission peaks, some of which
(if not all) harbour embedded sources. Her analysis does, thus, not necessarily
apply to general, cool IRDC material, but to gas that is influenced by embedded
protostars (see \S\ref{starformation}). This is reflected most directly
in the high-velocity outflows
seen in some sources, e.g, G11.11-0.12 and in the high kinetic
temperatures of order 40 to 60 K she derives for these.


\subsection{Morphology}\label{sec:morph}
A significant fraction of IRDCs (although no all) are
filamentary. Are IRDCs really filaments, i.e. elongated cylinders, or are they
sheets seen edge-on?
This is actually an important question linked to their evolutionary state.

Larson (1985), using theoretical arguments and numerical
simulations, argues that fragmentation is unlikely to occur
in an initially uniform cloud. Either an initial anisotropy or rotation
or a magnetic field will in general cause the cloud to collapse toward a
flattened or filamentary structure. Once overall collapse has been halted
and approximate equilibrium has been established, gravitational instability
can cause the resulting sheet or filament to break into fragments
of a characteristic mass that depends on the temperature and the surface density
of the cloud.

Larson's arguments are supported by the work of
Miyama et al. (1987a,b), who investigated the
fragmentation instability of an isothermal gas
layer, in order to see whether the observed
structures of many dense interstellar clouds are
the results of fragmentation of sheet like clouds.
Linear perturbation theory predicts fragmentation of
a parent sheet-like cloud in elongated structures,
and using further nonlinear analysis they found that, if fragments
are initially elongated, they become elongated more and more as they go
on collapsing, ending up as very slender cylinders, which fragment further.

The G11.11-0.12 IRDC has a distinct filamentary appearance (see Figs. 1 and 4).
Using a sophisticated computational technique Fiege et al. (2004)
compared observations to three different models of self-gravitating,
pressure-truncated filaments, namely the non-magnetic Ostriker (1964) model,
and two magnetic models from the literature. Analysing the $850 \mu$m SCUBA observations of
G11.11-0.12,
Johnstone et al. (2003) concluded that this source has a much steeper
$r^{-\alpha}, (\alpha \simgreat 3$) radial density profile than other (lower moss, lower extinction)
filaments, where the density varies approximately as $r^{-2}$,
This steep density profile is consistent
with the Ostriker model. After a wider search of parameter space, Fiege et al.
conclude that the observed radial structure of G11.11-0.12 can be understood in the
context of all three models.
Discrimination between the different models may be possible with
polarization measurements as the magnetic models predict
dominant poloidal magnetic fields that are
dynamically significant; G11.11-0.12 may be
radially supported by a poloidal field. Fiege et al. predict polarization patterns expected for
both magnetic models, which produce
different polarization patterns. Polarimetry  should, thus, be able to distinguish
between the two magnetic models or a non-magnetic model.

An instrument of choice will be PolKa, the Polarization Kamera
designed to be used together with the bolometer arrays developed at the
Max-Planck-Institut für Radioastronomie, for example with the
295 element $870\mu$m Large APEX BOlometer CAmera (LABOCA), which is soon to
be installed at the 12m Atacama Pathfinder EXperiment
telescope
(Siringo et al. 2004).

To decide the sheet or cylinder question, let us consider G11.11-0.12. That cloud
is at a distance of 3.6 kpc (Carey et al. 1998). The elongated
submillimeter emission has an extent of 24 pc in the long and an average $\sim 0.8$~pc
in the short axis. The density is uncertain: values between 10$^5$ and 10$^6$
\ccm\ have been derived by Carey et al.
(1998) from their \hhco\ data, Johnstone et al. (2003) argue for a few times $10^4$
\ccm.  The H$_2$ column density is between 0.2 and 2$\times 10^{22}$ cm$^{-2}$ (Carey et al. 2000).
Using the extremes of these values, we find that the extent along the line of sight
must be between a few times $10^{-3}$ and 0.5 pc, definitely ruling out a sheet seen edge on.

\subsection{Chemistry of IRDCs}\label{chemistry}
Complex organic (i.e. O- and C-bearing) molecules in the interstellar
medium are mostly found in hot, dense cores surrounded by high-mass protostars.
They frequently have very high over abundances
(factors of 100 -- 1000) compared to dark cloud values. These are usually explained by the evaporation of icy
dust grain mantles on which these molecules are formed in a cooler phase in the clouds lifetime
by hydrogenation of CO to H$_2$CO. Further hydrogenation leads to even more complex species.
(Relatively) complex molecules
are also found in cold dark clouds, with TMC-1 being a prominent nearby example
(See, e.g. Kaifu et al. 2004).
However, in the latter they all have very small abundances and are observable
only because of TMC-1's proximity (yielding a high filling factor) and its moderately
high density ($~10^4$ cm$^{-3}$) leading to substantial beam-averaged column densities.
This makes exotic (but not organic in the strict sense of the work) species
detectable, such as the polyyene carbon chains (Kaifu et al. 2004).


What is the organic content of normal molecular clouds? This, essentially,
is an unanswered question (the one example TMC-1 aside). Its answer
has profound impact on astrochemistry (are grain mantles really needed to
form these molecules?) and even astrobiology.
Their high column densities make IRDCs ideal laboratories
to address this question and potentially detect complex molecules. Such species
might be present and widespread
in many clouds, but would be rendered undetectable because of the
modest column densities of ordinary clouds:
Lines from almost all molecules significantly rarer than NH$_3$ or CH$_3$OH will
most likely be optically thin, which makes their line intensity directly
proportional to the column density.

Observations of molecules in IRDCs so far have concentrated on just a
few species: CO (several isotopomers),
H$_2$CO, CH$_3$OH, and NH$_3$. In addition, Teyssier et al. (2002)
observed several  HC$_3$N lines and two $k$-series of CH$_3$CCH.
The latter, a symmetric top, can be used as a temperature probe and its observations
yield values for the kinetic temperature, \Tkin, between 8 and 25 K; the higher
values found toward embedded objects. Large Veloity Gradient (LVG) model calculations
of HC$_3$N, $^{13}$CO, and C$^{18}$O yield densities larger than $10^5$ \ccm\ in the densest
parts. Teyssier et al. ascribe the relatively low observed intensities (a factor of a few
lower than in TMC-1) to a very low kinetic temperature (difficult to understand, as
TMC-1 is cold, too, $\approx 10$ K), a small filling factor or depletion on grains.

In Fig.\ 2, we show the spectra of different molecules observed towards
the brightest submm peak position of G11.11-0.12 (P1).

\label{fig:2}
\begin{figure}
\begin{tabular}{p{7cm}p{5.5cm}}
\includegraphics[height=0.73\linewidth,angle=-90]{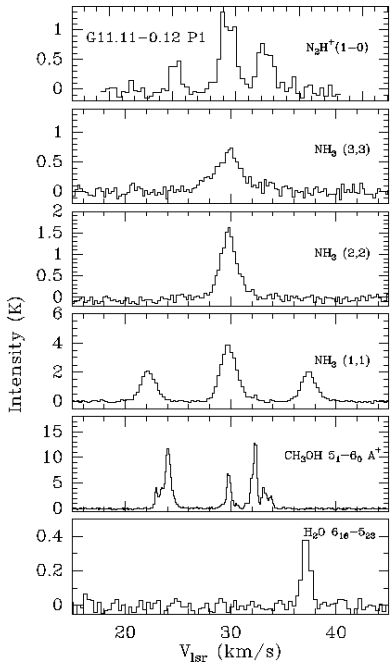}
&\caption{\textit{top to bottom}: Spectra of the N$_2$H$^{+}$ (1-0), NH$_3$ $(J,K) = (3,3)$, (2,2),
and (1,1) lines and the  CH$_3$OH 6.7~GHz and H$_2$O 22.2~GHz maser
transitions. All spectra were taken towards the submm peak
position P1 given by Carey et al. 2000 (see Table 1).  }\\
\end{tabular}
\end{figure}

The NH$_3$ and CH$_3$OH observations produced
interesting results: CH$_3$OH and NH$_3$
are overabundant by factors of 5 -- 10
relative to ``normal"(= lower density) and less turbulent dark clouds , such as TMC-1
(Leurini, dissertation and 2005, Pillai et al. 2005a in prep.).
In contrast, H$_2$CO is \textit{under}abundant by a factor of $\sim 50$
(Carey et al. 1998). Given this situation it is completely unclear
which molecules might be detectable and which ones not. Could it be that species
with emission sufficiently strong and widespread to be easily detectable
have until now been completely missed?

Systematic searches for other molecules will yield a more complete picture
of the chemistry of IRDCs, which, while  certainly  interesting in itself,
will also shed light on general formation mechanisms of complex molecules.
Moreover, they might help identify new temperature and density tracers and
allow studies of (molecule-)selective depletion.

\subsubsection{Ammonia}\label{ammonia}

To exploit \nhhh's properties as an
excellent molecular cloud thermometer (Danby et al. 1988)
10 IRDCs were studied
in the course of T. Pillai's dissertation (see also Pillai et al. 2005):
They were mapped in the
$(J,K) = $ (1,1) and (2,2) transitions near $1.3$~cm
wavelength ($\sim 23.7$~{GHz}) with the MPIfR Effelsberg 100m telescope.
The FWHM beam size at the frequencies of the \nhhh\ lines was $40''$. The
\nhhh\ emission correlates very well with MIR absorption and ammonia
peaks (in the following referred to as ``cores'') distinctly
coincide with dust continuum peaks, as shown in Fig.\ 3.
\label{fig:3}
\begin{figure}
\begin{center}
\includegraphics[height=0.73\linewidth,angle=-90]{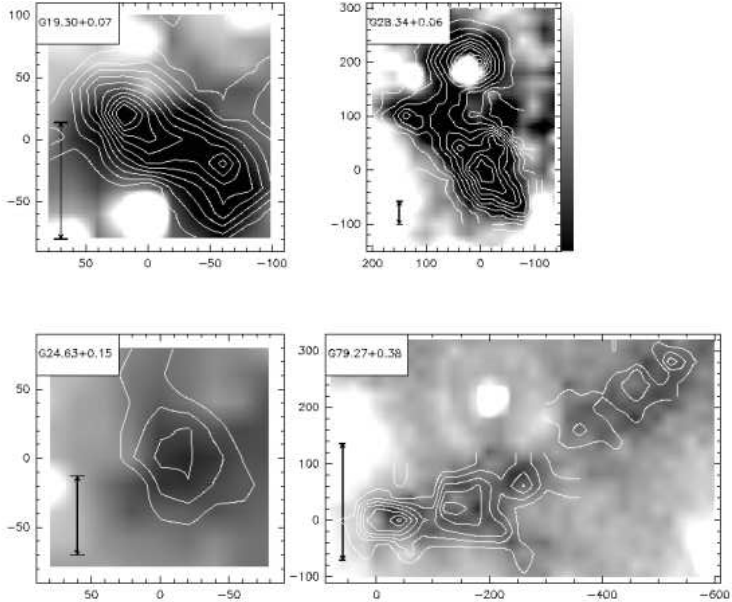}
\end{center}
 \caption{MSX images of the clouds at 8${\mu}m$ \textit{(greyscale)} with \nhhh\ (1,1)
 integrated intensity as contours. The contour levels are 2, 4, and 6
times the $1\sigma$ noise level. Tick marks are coordinate offsets
 (in arcseconds) relative to the positions given in Table 1 (from Pillai et al. 2005a).
 The bar marks a distance of 1 pc.}
\end{figure}

Several compact sources were detected within the clouds with sizes
smaller than the $\approx 40''$ FWHM beam size.  The total gas masses
derived for entire clouds from \nhhh\ data range from 10$^3$ -- 10$^4$
\Msun.

We can constrain the average gas temperatures to $10 \simless T \simless  20$~K.
The temperature distribution within clouds has also been analysed and we find significant
temperature gradients, with the temperature rising in outward direction,
in all of the cases where we have a good
signal-to-noise ratio throughout
the map. This outward rise in temperature we find in all except one core can be readily
interpreted as influence of the strong external UV field warming up the cloud.

The only case where we find a positive correlation between the gas temperature
and the integrated intensity is also the only case where the turbulence seems
to increase towards the core. This one has the most evolved core and is also the most
massive in our sample. We observe large line widths ($1 \le \Delta v \le 3.5$)~\kms,
hence turbulence plays an important role in the stability of this IRDC.
The column densities translate to extremely high $A_V$ of 55 -- $450$~mag,
therefore early star formation, if any, would be deeply embedded.
The virial parameter defined as $\alpha = {\frac{{5}{\sigma^{2}}{R}}{GM}}$
is $1.7\le\alpha\le4$ for most of the clumps. Hence the cores appear to be
unstable against gravitational collapse; in fact direct evidence for
collapse might be revealed from VLA observations we have recently obtained.

The fractional abundance of \nhhh\
(relative to $\rm H_2$) is $1-6\times 10^{-8}$. This together with the excellent
correlation in morphology of the dust and gas is consistent with the time
dependant chemical model for \nhhh\ of Bergin \& Langer (1997) and
implies that \nhhh\ remains undepleted. We can constrain the ages of IRDCs based
on this model to $\ge 10^7$ years for $\rm H_2$ densities $\ge 10^4$ \ccm,
assuming that the \nhhh\ has reached its chemical equilibrium abundance.
The time scales we derive for the clouds to disperse due to their own internal
motions, of a few Myrs, provide a better upper limit to the life time of these clouds.
 There are significant velocity gradients observed between the cores but we
 find that they are not attributable to rotation. The effects of external shock/outflow
 tracers need to be investigated.

Based on the observed line widths, the derived gas temperatures and the
\nhhh\ column densities, we made a comparison of IRDC
condensations with objects representing more evolved stages of high-mass
star formation like the High-mass Protostellar Objects (HMPOs) studied by
Sridharan et al. (2002) and Beuther et al. (2002). There is a clear trend
in temperature from the low
temperatures of the IRDCs to typical temperatures of 20 -- 30 K for the
HMPOs without (significant) HII region to the higher temperatures of
UCHII/hot core regions. The line widths in the HMPOs are generally
higher than those in the IRDCs.


\subsection{Magnetic fields}
Theoretical studies suggest that magnetic fields play a crucial role in the star
formation process. But contrary to other parameters like density, temperature,
the velocity field, and molecular abundances, it has been notoriously
difficult to determine $B$-fields in any regime of the interstellar medium (ISM)
from diffuse clouds to dense star-forming cores (see, e.g., Crutcher 1991 and these proceedings).

Virtually the only method for a direct determination of $B$ is the Zeeman effect,
which causes a frequency shift of the right-circularly polarized (RCP)
relative to the left circularly polarized (LCP) component of a spectral line from a molecule
with a suitable electronic structure and also from the 21 cm line of the hydrogen atom.

One of the few interstellar molecules with detectable Zeeman splitting
is hydroxyl (OH), whose ground-state hyperfine structure (hfs)
transitions near 18 cm wavelength (at 1665, 1667, 1612, and 1720 MHz)
have measurable splittings.
OH is found in the general molecular interstellar medium and can be
detected in clouds with densities $\ge$ a few $10^3$
cm$^{-3}$. However, it is also found, at elevated abundance, in the
dense, expanding envelopes of ultracompact HII regions, which have
densities $\ge 10^7$ \ccm\ (Hartquist et al. 1995).

$B$ fields of order a few tens of $\mu$G have been found from OH Zeeman measurements
of low density  dark clouds (see, e.g. Goodman et al. 1989; Crutcher \&\ Troland 2000),
while much stronger, few mG, fields are derived for the much denser maser regions
(see, e.g., Fish et al. 2003).

While it is relatively easy to measure Zeeman-splitting in OH masers, over the years, large
amounts of observing time have been dedicated to measuring Zeeman-splitting in lower density
clouds with few, but precious results. A picture has emerged in which the $B$-field strength
increases with density, $n$, i.e. $|B| \propto n^\kappa$. From theoretical
arguments (conservation of magnetic flux and mass) one expects $\kappa = 2/3$ for a
collapsing cloud if the $B$-field is unimportant throughout the collapse and $\kappa =1/2$
in the opposite case (Crutcher 1991). Models of ambipolar diffusion-driven
cloud contraction deliver $\kappa \approx 0.47$ (Fiedler \&\ Mouschovias 1993).

See Fig. 1 of from Padoan \& Nordlund 1999  for a recent compilation
of measured $B$-field strengths vs. density.
It is apparent from this figure that there is a dearth
of $B$-field data points for densities between $10^5$ and $10^6$
\ccm.  $B$-field measurements of IRDCs will probe just this
highly interesting portion of
parameter space.

A preliminary survey with the Effelsberg 100m telescope showed
OH absorption in both the 1665 and the 1667 MHz hyperfine lines
with total (Stokes $I$) intensities of order $-1$ K or deeper
in several IRDCs.
Given the IRDCs' densities cited above we would
expect $B$ of order several hundred mG, similar to the values found in
high-mass star-forming regions. A 2 hour integration on one source,
G yielded an upper limit
of 135~$\mu$G at a 99\%
(3$\sigma$) confidence level, consistent with the upper
limits derived by  Crutcher at al. (1993) for regions of similar density.

The $B$-field morphology will be determined from submillimeter polarization
observations (see \ref{sec:morph}). Feldman et al. (2003) report in an abstract
SCUBA polarization observations
of MSX IRDCs where they quote very high percentage polarizations ($\sim6$\%)
and find that inferred magnetic field directions are
correlated with the cloud structure. There seems to
be trend for $B$ to align along the direction
of a filament. Bright, compact sources in the filaments are
much less polarized, and their inferred $B$-field
directions are perpendicular to the orientation of the filaments.

\section{Ongoing star formation in IRDCs}\label{starformation}
While large volumes of IRDCs appear to be devoid of signposts
of ongoing star formation, such as ultracompact HII regions
and/or CH$_3$OH, OH or H$_2$O masers, isolated centers of high or intermediate star
formation are found in many clouds.

Teyssier et al. (2002) found that OH and class II CH$_3$OH masers et
al. (1995) are associated with positions of (not overly pronounced)
peak emission from the column density tracer C$^{18}$O in the IRDCs
DF+9.86-0.04 and DF+30.23-0.20. These are close to dust emission
peaks. Since CH$_3$OH masers are unambiguous tracers of high-mass star
formation, we have obtained data on the 6.7 GHz CH$_3$OH maser
transition, towards a sample of $\sim$~50 dark clouds with a high ($>$
25\%) rate of detections.

Maybe to date the best-studied example of
a star-forming core in an IRDC is the 850 $\mu$m emission
peak in G11.11-0.12 studied in detail by Pillai et al. (2005a; in press).
Coincident with a compact dust continuum source are
both, an H$_2$O and a CH$_3$OH maser as shown in the inset of Fig.\ 4.
Interferometric imaging with the
Australia Telescope Compact Array show the CH$_3$OH emission, which has a
total velocity spread of $\approx 11$ \kms\ to have a velocity
gradient with emission at different velocities aligned in a
line, reminiscent of a disk. Other persuasive arguments
for an embedded source are the detection of emission in the high
excitation (3,3) line of ammonia with a wider linewidth than the lower
excitation (1,1) and (2,2) lines (see Fig. 2). Model fits to all three
\nhhh\ lines indicate a compact source with a size of $\approx 3''$, characterized
by a rotation temperature of 60 K, while the more extended emission
from the ambient cloud has a rotation temperature of 15 K. The NH$_3$ column density
of the hot, compact component is 9 times higher than that of the
cool extended one. Finally, the infrared spectral energy distribution
is best modelled by a source with a luminosity of 1200 \Lsun, corresponding to
a ZAMS star of mass 8 \Msun.  $K_s$-band 2MASS data show what possibly is reflected light
emanating from the protostellar source, which is embedded in a compact
mm-wavelength dust continuum source imaged with the Berkeley-Illinois-Maryland Array (BIMA).

\label{fig:4}
\begin{figure*}
\begin{center}
\includegraphics[height=\linewidth,angle=-90]{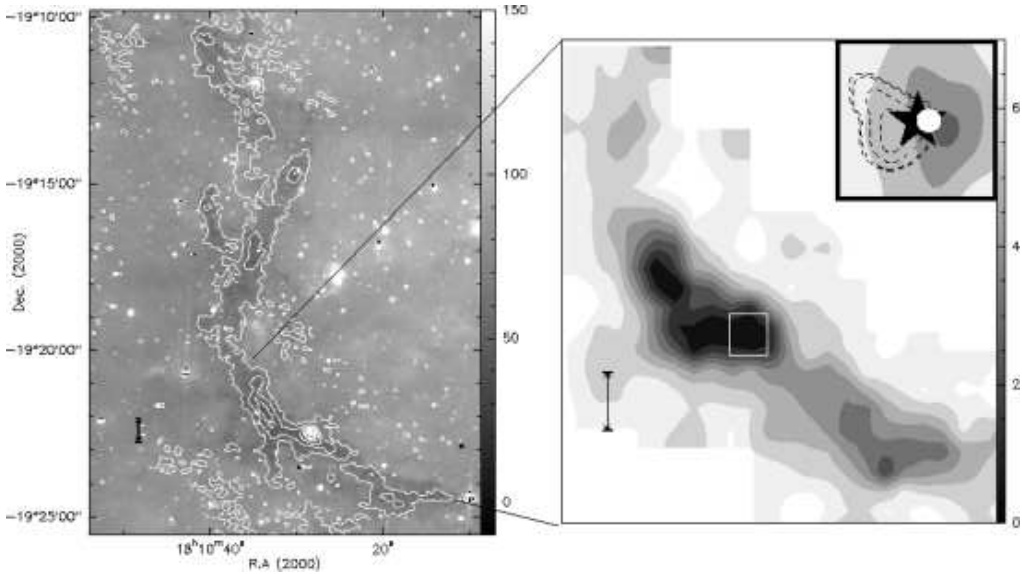}
\end{center}
 \caption{\textit{Left:} The $8 ~ \rm \mu m$ image of G11.11-0.12 with the
 SCUBA $850 ~ \rm \mu m$ map (Carey et al.\ 2000) overlaid. \textit{Right:}
 Southern filament of the SCUBA map in grey scale. The bar marks a distance of 1 pc.
 The \textit{square}
 delineates the position of the active star formation site P1, details
 of which are shown in the right upper corner inset.  Here the
 \textit{greyscale} shows a BIMA 3~mm continuum image and a 2MASS $K_s$ band image
 in shown in dashed contours.  The \textit{star} denotes the H$_2$O maser
 position and the \textit{filled circle} the CH$_3$OH maser position (from Pillai et al. 2005b).}
\end{figure*}

\section{Conclusions}\label{sec:concl}
The earliest phases of high-mass star formation are expected to
be massive (a few hundred to thousand M$_\odot$), cold (10 -- 20 K) and quiescent clouds,
emitting primarily at  (sub)mm  wavelenghts and containing
no obvious IR sources or star formation tracers.
One approach to identify the earliest, cold phases of massive star
formation is to search for objects which appear in absorption at MIR wavelengths.
Thus IRDCs are the most potential candidates for studying these initial conditions.
Centimeter through submm observations reveal that typical IRDCs have gas densities
$>  10^{6} ~ {\rm cm}^{-3}$, temperatures $< 20 ~{\rm K}$ and sizes
of 1 -- 10~pc but studies of their star formation content are still rare.

Studies up to now seem to show that they are not all cold and quiescent.
IRDCs appear to harbour sources of different evolutionary stages,
not all of them necessarily in the high-mass regime. A better classification
scheme based on molecular gas content, MIR contrast and extend is needed to
compare IRDCs with local molecular cloud complexes (not clouds).
Extensive studies of their energetics, kinematics and chemistry are essential
to ascertain their role in forming stars, massive or otherwise. These would be the
ideal test grounds for testing the present theories of forming massive stars via
turbulent cores (McKee \& Tan 2003; see also their contributions to these proceedings).
We will need large, Galaxy wide surveys
to understand the formation of IRDCs and their lifetimes.

We would like to thank Malcolm Walmsley for comments on the manuscript.


\begin{thebibliography}{}

\bibitem[Bergin and Langer \ (1997)]{Ber97} Bergin, E.A. and Langer, W.D.1997,
\textit{ApJ}, 486,316

\bibitem[Beuther et al.(2002)]{2002ApJ...566..945B} Beuther, H., Schilke,
P., Menten, K.~M., Motte, F., Sridharan, T.~K., \& Wyrowski, F.\ 2002,
\textit{ApJ}, 566, 945


\bibitem[Bohlin et al.(1978)]{1978ApJ...224..132B}
        {Bohlin, R.~C., Savage, B.~D., \& Drake, J.~F.} 1978,
        \textit{ApJ} 224, 132


\bibitem[Carey et al.(1998)]{1998ApJ...508..721C}
        {Carey, S.~J., Clark,
        F.~O., Egan, M.~P., Price, S.~D., Shipman, R.~F., \& Kuchar, T.~A.} 1998,
        \textit{ApJ} 508, 721

\bibitem[Carey et al.(2000)]{2000ApJ...543L.157C}
        {Carey, S.~J., Feldman, P.~A., Redman, R.~O., Egan, M.~P.,
        MacLeod, J.~M., \& Price, S.~D.} 2000,
        \textit{ApJ} (Letters) 543, L157

\bibitem[Crutcher(1991)]{1991IAUS..147...61C} Crutcher, R.~M.\ 1991, IAU
Symp.~147: Fragmentation of Molecular Clouds and Star Formation, 147, 61



\bibitem[Crutcher et al.(1993)]{1993ApJ...407..175C} Crutcher, R.~M.,
Troland, T.~H., Goodman, A.~A., Heiles, C., Kazes, I., \& Myers, P.~C.\
1993, \textit{ApJ}, 407, 175

\bibitem[Crutcher \& Troland(2000)]{2000ApJ...537L.139C} Crutcher, R.~M.,
\& Troland, T.~H.\ 2000, \textit{ApJ}, 537, L139

\bibitem[Danby et al.(1988)]{1988MNRAS.235..229D} Danby, G., Flower, D.~R.,
Valiron, P., Schilke, P., \& Walmsley, C.~M.\ 1988,
\textit{MNRAS}, 235, 229


\bibitem[Draine \& Lee(1984)]{1984ApJ...285...89D} Draine, B.~T., \& Lee,
H.~M.\ 1984,
\textit{ApJ}, 285, 89


\bibitem[Egan et al.(1998)]{1998ApJ...494L.199E} Egan, M.~P., Shipman,
R.~F., Price, S.~D., Carey, S.~J., Clark, F.~O., \& Cohen, M.\ 1998, \textit{ApJ},
494, L199


\bibitem[Feldman et al.(2003)]{2003cdsf.conf..292F} Feldman, P.~A., Redman,
R.~O., Carey, S.~J., \& Wyrowski, F.\ 2003, SFChem 2002, eds.  C. L.~Curry \&\
Michel Fich, NRC Press, Ottawa, Canada, p. 292.

\bibitem[Fiedler \& Mouschovias(1993)]{1993ApJ...415..680F} Fiedler, R.~A.,
\& Mouschovias, T.~C.\ 1993, \textit{ApJ}, 415, 680

\bibitem[Fiege et al.(2004)]{2004ApJ...616..925F} Fiege, J.~D., Johnstone,
D., Redman, R.~O., \& Feldman, P.~A.\ 2004,
\textit{ApJ}, 616, 925

\bibitem[Fish et al.(2003)]{2003ApJ...596..328F} Fish, V.~L., Reid, M.~J.,
Argon, A.~L., \& Menten, K.~M.\ 2003, \textit{ApJ}, 596, 328

\bibitem[Garay et al.(2004)]{2004ApJ...610..313G} Garay, G., Fa{\' u}ndez,
S., Mardones, D., Bronfman, L., Chini, R., \& Nyman, L.\ 2004, \textit{ApJ}, 610,
313

\bibitem[Goodman et al.(1989)]{1989ApJ...338L..61G} Goodman, A.~A.,
Crutcher, R.~M., Heiles, C., Myers, P.~C., \& Troland, T.~H.\ 1989, \textit{ApJ},
338, L61


\bibitem[Hartquist et al.(1995)]{1995MNRAS.272..184H} Hartquist, T.~W.,
Menten, K.~M., Lepp, S., \& Dalgarno, A.\ 1995,
\textit{MNRAS}, 272, 184


\bibitem[Hennebelle et al.(2001)]{2001A&A...365..598H}
        Hennebelle, P., P{\'e}rault, M., Teyssier, D., \& Ganesh, S.\ 2001,
        \textit{A\&A} 365, 598

\bibitem[Johnstone et al.(2003)]{2003ApJ...588L..37J} Johnstone, D., Fiege,
J.~D., Redman, R.~O., Feldman, P.~A., \& Carey, S.~J.\ 2003, \textit{ApJ}, 588,
L37

\bibitem[Kaifu et al.(2004)]{2004PASJ...56...69K} Kaifu, N., et al.\ 2004,
\textit{PASJ}, 56, 69


\bibitem[Larson(1985)]{1985MNRAS.214..379L}
        {Larson, R.~B.} 1985,
        \textit{MNRAS}, 214, 379

\bibitem[Leurini et al.(2004)]{2004A&A...422..573L} Leurini, S., Schilke,
P., Menten, K.~M., Flower, D.~R., Pottage, J.~T., \& Xu, L.-H.\ 2004,
\textit{A\&A}, 422, 573

\bibitem[Lis et al.(1994)]{1994ApJ...423L..39L} Lis, D.~C., Menten, K.~M.,
Serabyn, E., \& Zylka, R.\ 1994, \textit{ApJ}, 423, L39

\bibitem[Lis \& Carlstrom(1994)]{1994ApJ...424..189L} Lis, D.~C., \&
Carlstrom, J.~E.\ 1994, \textit{ApJ}, 424, 189

\bibitem[Lis \& Menten(1998)]{1998ApJ...507..794L} Lis, D.~C., \& Menten,
K.~M.\ 1998, \textit{ApJ}, 507, 794



\bibitem[Lutz et al.(1996)]{1996A&A...315L.269L}
        {Lutz, D., et al.} 1996, \textit{A\&A}, 315, L269


\bibitem[Mangum \& Wootten(1993)]{1993ApJS...89..123M}
        {Mangum, J.~G., \& Wootten, A.} 1993,
        \textit{ApJS}, 89, 123

\bibitem[McKee \& Tan(2003)]{2003ApJ...585..850M} McKee, C.~F., \& Tan,
J.~C.\ 2003, \textit{ApJ}, 585, 850


\bibitem[Mezger et al.(1990)]{1990A&A...228...95M}
        {Mezger, P.~G., Zylka, R., \& Wink, J.~E.} 1990,
        \textit{A\&A}, 228, 95

\bibitem[Mezger et al.(1987)]{1987A&A...182..127M}
        {Mezger, P.~G., Chini, R., Kreysa, E., \& Wink, J.} 1987,
        \textit{A\&A}, 182, 127


\bibitem[Miyama et al.(1987)]{1987PThPh..78.1051M}
        {Miyama, S.~M., Narita, S., \& Hayashi, C.} 1987a,
        \textit{Progress of Theoretical Physics}, 78, 1051

\bibitem[Miyama et al.(1987)]{1987PThPh..78.1273M}
        {Miyama, S.~M., Narita, S., \& Hayashi, C.} 1987b,
        \textit{Progress of Theoretical Physics}, 78, 1273


\bibitem[Mundy et al.(1987)]{1987ApJ...318..392M}
        {Mundy, L.~G., Evans, N.~J., Snell, R.~L., \& Goldsmith, P.~F.} 1987,
        \textit{ApJ} 318, 392

\bibitem[Ostriker(1964)]{1964ApJ...140.1529O} Ostriker, J.\ 1964, \textit{ApJ},
140, 1529

\bibitem[Padoan \& Nordlund(1999)]{1999ApJ...526..279P} Padoan, P., \&
Nordlund, {\AA}.\ 1999, \textit{ApJ}, 526, 279


\bibitem[]{}  Pillai, T ., Wyrowski, F,  Menten, K.~M., \&\ Kr\"ugel, E., \textit{A\&A},
2005a, in press

\bibitem[]{} Pillai, T ., Wyrowski, F,  Menten, K.~M., \&\ Carey, S.~J., \textit{A\&A},
2005b, to be submitted

\bibitem[Price(1995)]{1995SSRv...74...81P}
        {Price, S.~D.} 1995,
        \textit{Space Science Reviews} 74, 81

\bibitem[Perault et al.(1996)]{1996A&A...315L.165P}
        Perault, M., et al.\ 1996,
        \textit{A\&A} 315, L165


\bibitem[Price et al.(2001)]{2001AJ....121.2819P}
        {Price, S.~D., Egan, M.~P., Carey, S.~J., Mizuno, D.~R.,
        \& Kuchar, T.~A.} 2001,
        \textit{AJ} 121, 2819

\bibitem[Redman et al.(2003)]{2003ApJ...586.1127R}
        {Redman, R.~O., Feldman, P.~A., Wyrowski, F., C{\^ o}t{\' e}, S.,
        Carey, S.~J., \& Egan, M.~P.} 2003,
        \textit{ApJ} 586, 1127


\bibitem[Simon et al.(2004)]{2004ASPC..317..159S} Simon, R., Shah, R.~Y.,
Rathborne, J., Jackson, J.~M., Bania, T.~M., Clemens, D.~P., \& Heyer,
M.~H.\ 2004, ASP Conf.~Ser.~317: Milky Way Surveys: The Structure and
Evolution of our Galaxy, 317, 159


\bibitem[Sridharan et al.(2002)]{2002ApJ...566..931S} Sridharan, T.~K.,
Beuther, H., Schilke, P., Menten, K.~M., \& Wyrowski, F.\ 2002,
\textit{ApJ}, 566,
931


\bibitem[Teyssier,Hennebelle, \& P\'erault]{2002A&A...382..624T}
Teyssier, D., Hennebelle, P., \& P\'erault, M.\ 2002,
\textit{A\&A}, 382, 624










\end{thebibliography}
\end{document}